\pdfoutput=1
\documentclass[preprint, tightenlines]{revtex4-1}
\usepackage{verbatim}
\usepackage{amsmath}
\usepackage{amssymb}

\def\beq{\begin{equation}}
\def\eeq{\end{equation}}
\def\bea{\begin{eqnarray}}
\def\eea{\end{eqnarray}}

\newcommand{\ket}[1]{\left|{#1}\right>}
\newcommand{\bra}[1]{\left<{#1}\right|}
\newcommand{\paren}[1]{\left({#1}\right)}

\newcommand{\absol}[1]{\left|{#1}\right|}

\renewcommand{\square}[1]{\left[{#1}\right]}
\newcommand{\vect}[1]{\mathbf{#1}}

\usepackage{graphicx}
\begin{document}

%% NOTE: TITLE PAGE & TOC NOT USED FOR MANUSCRIPT SUBMISSIONS %%
%\title{Template and style guide for authors submitting to \textit{Optics Express}}

%\vskip4pc

%\tableofcontents
%\clearpage
%% NO TITLE PAGE FOR OPEX SUBMISSIONS %%

%% START HERE
%%%%%%%%%%%%%%%%%% title page information %%%%%%%%%%%%%%%%%%
\title{Coherent optical non-reciprocity in axisymmetric resonators}

\author{Erik J. Lenferink, Guohua Wei, Nathaniel P. Stern}

\affiliation{Department of Physics and Astronomy, Northwestern University, 2145 Sheridan Road, Evanston, IL 60208, USA}

\email{n-stern@northwestern.edu} %% email address is required



\begin{abstract}
We describe an approach to optical non-reciprocity that exploits the local helicity of evanescent electric fields in axisymmetric resonators. By interfacing an optical cavity to helicity-sensitive transitions, such as Zeeman levels in a quantum dot, light transmission through a waveguide becomes direction-dependent when the state degeneracy is lifted.  Using a linearized quantum master equation, we analyze the configurations that exhibit non-reciprocity, and we show that reasonable parameters from existing cavity QED experiments are sufficient to demonstrate a coherent non-reciprocal optical isolator operating  at the level of a single photon.
\end{abstract}

%%%%%%%%%%%%%%%%%%%%%%% References %%%%%%%%%%%%%%%%%%%%%%%%%

\maketitle

%%%%%%%%%%%%%%%%%%%%%%%%%%  body  %%%%%%%%%%%%%%%%%%%%%%%%%%
\section{Introduction}

Under most circumstances, Maxwell's equations of electromagnetism suggest that the propagation of light obeys a principle of reciprocity: transmission from point $A$ to point $B$ is the same as that from $B$ to $A$~\cite{Potton2004}. In many cases breaking of this directional symmetry is desirable, most notably to protect sensitive optical elements from scattered light (optical isolators) or to implement direction-dependent logic (circulators). Increasing focus on all-optical fiber networks and photonic integrated circuits~\cite{Politi2008, Politi2009, Shadbolt2011,Obrien2009} for replacing electronic circuits for low-power opto-electronic networks~\cite{Kilper2012} highlights the need for new implementations of non-reciprocal devices such as optical isolators and optical circulators in the context of integrated photonics. Furthermore, applications for non-reciprocal behavior have recently emerged in quantum information processing and many-body physics simulation using photons.  In this context, direction-dependent phase shifts at the single photon level~\cite{Koch2010, Wang2008, Haldane2008, Hafezi2011, Hafezi2012} are important components for simulation of quantum Hall states with light, suggesting exciting applications for non-reciprocal elements in quantum optical networks.

The promise of both classical and quantum photonics~\cite{Obrien2009} drives exploration of new mechanisms for optical non-reciprocity at micron and sub-micron scales in which control of light in wavelength-scale environments with high device densities can be achieved. Traditional non-reciprocal optics exploits magnetic materials~\cite{Bi2011} and magneto-optical phenomena to break directional symmetry.  Despite recent research exploring alternative realizations based on diverse mechanisms such as cavity nonlinearities~\cite{Soljacic2003, Fan2012}, mode conversion~\cite{Feng2011}, spin polarization~\cite{Trowbridge2011}, coupled atoms and quantum dots~\cite{Shen2011, Mi2011}, and optomechanics~\cite{Manipatruni2009, Hafezi2012}, widespread implementation of non-reciprocity in integrated micro- and nano- photonics remains challenging.
Major hurdles include shrinking the magnetic or nonlinear medium to the micro- or nano-scale useful for integrated optics~\cite{Potton2004} and reducing dependence on power, which can limit isolation performance in the single photon quantum regime~\cite{Hafezi2012}.  Satisfying these demands requires novel approaches to breaking directional symmetry for non-reciprocal integrated photonics.

Here, we investigate non-reciprocal optical transmission in a hybrid device consisting of a spin-split quantum dot (QD) located in the evanescent field of a whispering gallery mode (WGM) microresonator in a regime of coherent light-matter interactions. We argue that direction-dependent, non-reciprocal behavior can manifest in an axisymmetric geometry from the local helicity of the electric field in the evanescent region where an emitter couples to light~\cite{Lacroute2012,Goban2012, Junge2013}. Extending the approach for a linear waveguide in~\cite{Shen2011}, we show that inducing a spin splitting in a V-type three-level system can achieve optical isolation using a coherent dynamical model in a low-power limit~\cite{Hafezi2012,Aoki2006, Alton2011, Junge2013}. This approach demonstrates how to exploit the local helicity of evanescent electric fields for optical non-reciprocity of coherent photons, similar to recent approaches using nonlinear optomechanical coupling~\cite{Hafezi2012}.

\section{Model and Approach}

\begin{figure}[tb]
	\centering
	\includegraphics[scale=1.00]{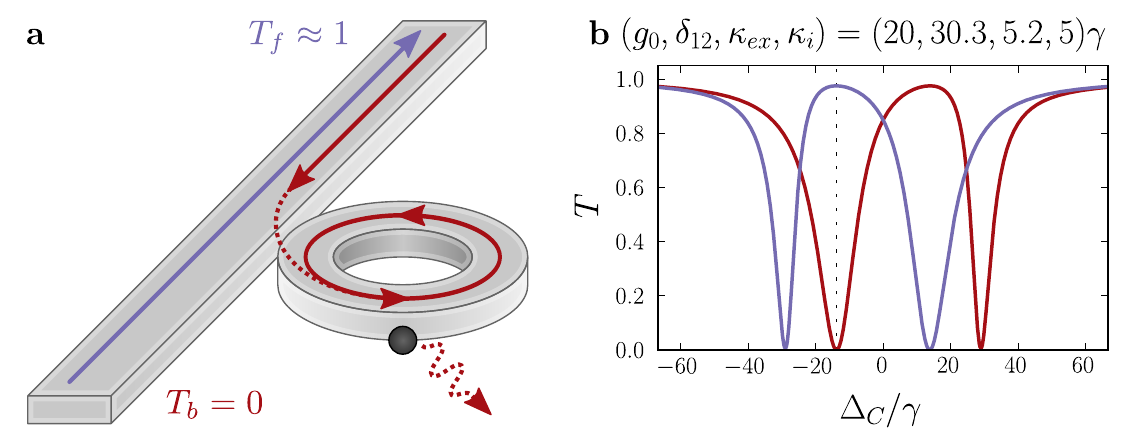}
	\caption{(a) The proposed nonreciprocal device acting as an optical diode. (b) Forward (blue) and backward (red) transmission spectra for a nonzero QD excited state splitting. Coupling between the counter-propagating resonator modes and the helicity-sensitive Zeeman-split excited states of the QD results in directional asymmetry of the spectra.  At a certain frequency (shown by the dashed line) the system exhibits a high degree of contrast $T_{\rm f}/T_{\rm b}$.}
	\label{3dring}
\end{figure}

We consider a system of an axisymmetric microcavity evanescently coupled to a V-type three-level single quantum dot (QD)~\cite{Shen2011, Mi2011}. The microcavity is evanescently coupled to a waveguide so that propagating waves from one direction predominantly traverse the resonator clockwise and those from the other direction couple to counter-clockwise propagating modes. The microcavity mediates an interaction between this input field and the QD exciton that can depend on the \emph{direction} of light propagation through the photonic system.  For guided modes in such a microcavity, the electric fields of the two counter-propagating modes possess opposite rotational helicity at each location.  This helicity can drive polarization-sensitive optical excitations, as has been demonstrated in atomic cavity QED and atom trapping~\cite{Junge2013, Lacroute2012, Goban2012}. By inducing an energy splitting between the two polarization-sensitive QD excited states, i.e. with a magnetic field, non-reciprocal transmission through the optical system can be realized. A similar idea exploiting the helicity of the electric field of waveguides was explored in a proposal by Shen \textit{et al}.~\cite{Shen2011}. In contrast to that approach, which assumed a distinct single photon propagating in the waveguide, our model treats a coherent input field, which can be useful for practical modeling of the dynamics that would be observed in a standard photon counting measurement.

The organization of the paper is as follows. First, we consider the local rotational helicity in the evanescent field of counter-propagating modes in axisymmetric resonators. This result is then used to deduce an effective Hamiltonian for a cavity-QD system, from which transmission spectra of the photonic system are found using a master equation approach.  By lifting the degeneracy of the two excited states, optical non-reciprocity will be shown to occur, and the conditions for achieving optical non-reciprocity are analyzed. Finally, we consider a possible cavity QED system in which these parameters for directional asymmetry can be realized.

\section{Evanescent Helicity in Whispering Gallery Modes}

\begin{figure}[tb]
	\centering
    \includegraphics[scale=1.0]{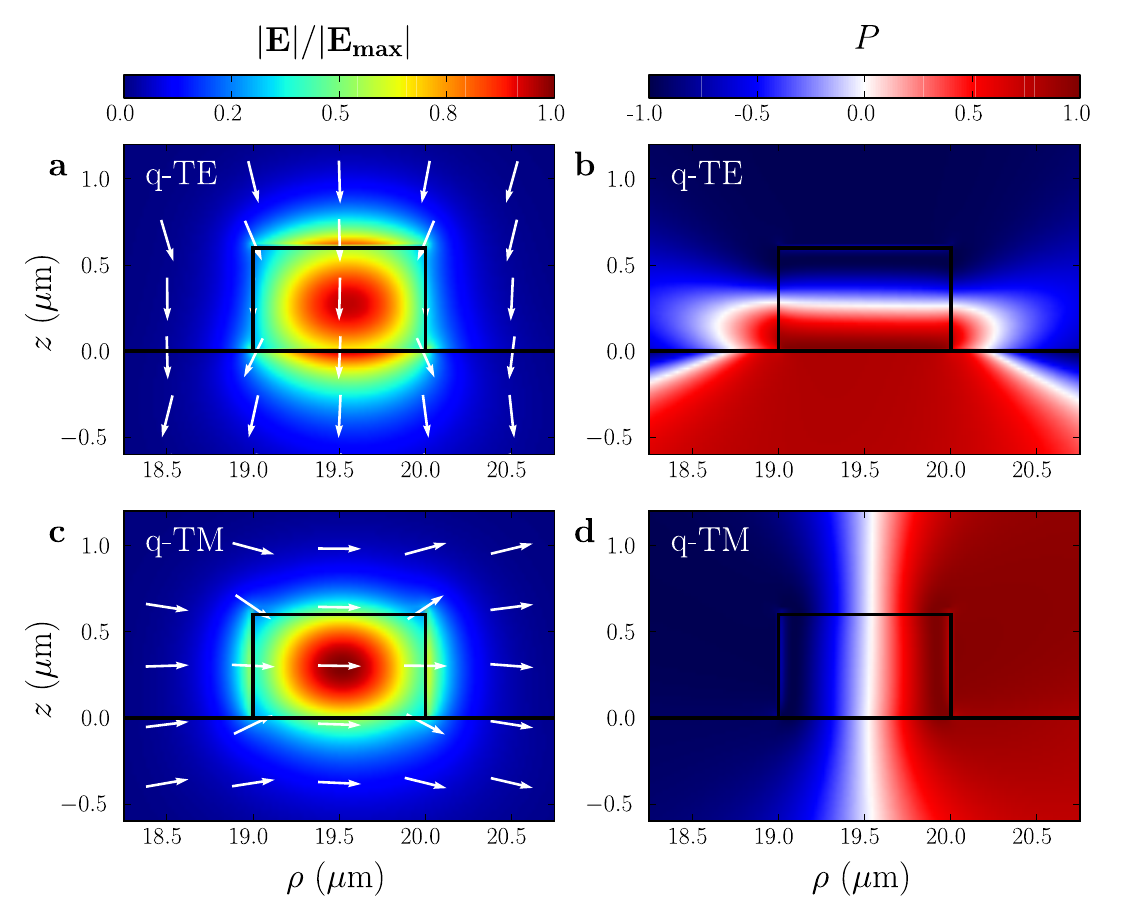}
	\caption{Cross-sectional plots of normalized $|\mathbf{E}|$ and $P$ for the $M=129$ CW fundamental (a-b) quasi-TE and (c-d) quasi-TM modes of a silicon nitride microring of outer radius~20~$\mu$m. Arrows indicate the direction of the $\vect{E}$ field in the $\rho$-$z$ plane. The frequency of this particular mode is $\omega_{\rm C}/2\pi = 194.0$ THz, although the general features are independent of $M$, geometry, and frequency. The region of the evanescent field directly (a-b) on top or (c-d) on the side of the cavity has a value of $P$ near unity and a large value of $\absol{\mathbf{E}}$. Finite element calculations were performed with COMSOL Multiphysics \cite{Oxborrow2007}.}
	\label{epplots}
\end{figure}

Axisymmetric microcavities support a degenerate pair of propagating modes for every resonant frequency $\omega_C$, one propagating clockwise (CW) and the other counter-clockwise (CCW). These modes are labelled by integer-valued azimuthal mode numbers of the same magnitude $M$ but opposite signs: $M > 0$ for CW and $M<0$ for CCW in our geometry~(Fig.~\ref{system}a). Working in cylindrical coordinates and omitting the overall time dependence $e^{-i \omega_C t}$, the electric field solution of a CW WGM of can be expressed as~\cite{Snyder1983,Shen2009}
\bea
	\mathbf{E}_{\rm CW}(\mathbf{r}) &=& \mathbf{E}_M(\rho,z)e^{i M \phi } \nonumber\\
&=& \paren{ E_\rho, i E_\phi, E_z } e^{i M \phi}
	\label{ecw}
\eea
where $\mathbf{E}_M(\rho,z)$ is a mode cross-section function, and $E_\rho$, $E_\phi$, and $E_z$ are real functions of $\rho$ and $z$. In general, both transverse ($\rho$ and $z$) and longitudinal ($\phi$) components are permitted, with a $\pi/2$ phase difference between them. Consequently, at a fixed coordinate location, the real electric field rotates in time; $\vect{E}$ is not linearly polarized but rather possesses rotational helicity. This is analogous to a traditional circularly polarized electric field that rotates around a helicity axis $\hat{e}_\odot$ parallel to the direction of propagation. In contrast, for an axisymmetric WGM, the axis $\hat{e}_\odot$ is perpendicular to the direction of propagation. Both the magnitude and axis of the helicity vary spatially, depending on the relative values $E_\rho$, $E_\phi$, and $E_z$ in the mode.  The axis is also different between quasi-transverse electric (TE) and quasi-transverse magnetic (TM) modes~(Fig.~\ref{epplots}).

The electric field of the corresponding CCW WGM is
\bea
	\mathbf{E}_{CCW}(\mathbf{r}) &=& \mathbf{E}_{-M}(\rho,z)e^{-i M \phi } \nonumber\\
&=& \paren{ E_\rho, -i E_\phi, E_z } e^{-i M \phi}
	\label{eccw}
\eea
where $E_\rho$, $E_\phi$, and $E_z$ are the same as in Eq.~\ref{ecw}. Whereas the mode spatial profile of $\mathbf{E}_{\rm CW}$ and $\mathbf{E}_{\rm CCW}$ are identical by symmetry, the phase of the longitudinal $\phi$ component has opposite sign for the two modes due to their contrasting direction of propagation. Consequently, the fields possess opposite helicity: the real part of $\vect{E}_{\rm CCW}$ rotates counter to that of $\vect{E}_{\rm CW}$.  An interaction sensitive to this electric field helicity will depend on the direction of the propagating light.

We quantify the degree of helicity by first defining a position-dependent unit vector $\hat{e}_\perp (\rho, z)$ perpendicular to $\hat{\phi}$ and parallel to the transverse electric field components. The basis vectors for positive (+1) and negative (-1) helicity can be written
\begin{equation}
	\hat{e}_\pm (\rho, z) = ( \hat{e}_\perp \pm i \hat{\phi} ) / 2^{1/2}
\end{equation}
The helicity quantization axis perpendicular to the plane in which the local field rotates can be defined as $\hat{e}_\odot(\rho, z) = \hat{e}_\perp \times \hat{\phi}$.  These unit vectors are spatially-dependent since the ratio of $E_\rho$ to $E_z$ varies throughout the mode. For a particular WGM, $\vect{E}_M$ is generally nearly parallel to $\hat{\rho}$ (quasi-TM) or to $\hat{z}$ (quasi-TE) in the evanescent region; the direction $\hat{e}_\odot(\rho, z)$ does not change appreciably for a particular mode profile. The degree of helicity at each position $\vect{r}$ can be written in terms of the helicity components along the local axis $\hat{e}_\odot(\rho, z)$, denoted $E_\pm(\rho, z) = \mathbf{E} \cdot \hat{e}_\pm(\rho, z)$:
\begin{equation}
	P(\rho,z) = ( | E_{+} |^2 - | E_{-} |^2 ) / | \mathbf{E}|^2
\end{equation}
$P$ ranges from $-1$, corresponding to a field of perfect negative helicity, to $+1$, corresponding to perfect positive helicity. Fig.~\ref{epplots} shows finite-element calculations of the mode profile and helicity for representative TE and TM ring resonator modes.

Because of the symmetries underlying Eqs.~\ref{ecw} and~\ref{eccw}, the $P$ values for a pair of counter-propagating WGMs have equal magnitude but opposite sign at each position. Consequently, just one function $P(\rho,z)$ is necessary to describe the mode pair. If we denote the value of $P$ for the CW mode as $p$, then the $P$ value for the CCW mode is $-p$ at the same position. The components $E_\pm$ for the two modes can be written as a functions of $p$:
\begin{align}
	E_{\rm CW,\pm}(\vect{r}) &= \absol{ \mathbf{E}_M(\rho,z) } e^{i M \phi} \square{(1 \pm p(\rho,z) )/2}^{1/2}
	\label{posnegcw}\\
	E_{\rm CCW,\pm}(\vect{r}) &= \absol{ \mathbf{E}_M(\rho,z) } e^{-i M \phi} \square{(1 \mp p(\rho,z) )/2}^{1/2}
	\label{posnegccw}
\end{align}
Importantly, $\vect{E}$ is projected onto a \emph{local} helicity basis with a spatially-varying rotation axis $\hat{e}_\odot$.

Although the electric field of a WGM possesses local helicity, it is not circularly polarized in the conventional sense. For circularly polarized plane waves, both real and imaginary components of the electric and magnetic fields are transverse to the propagation direction. For a WGM, the helicity axis $\hat{e}_\odot$ is transverse to the propagation direction $\hat\phi$. Despite this distinction, in the dipole approximation, a field $\vect{E}$ with large $p$ can excite the same optical transitions as circularly polarization.  This phenomenon has been observed experimentally for an axisymmetric bottle resonator, in which different Zeeman sub-levels of a Rubidium atom participate in cavity QED~\cite{Junge2013};  it has also been inferred from the linewidth broadening of atoms in nanofiber traps~\cite{Lacroute2012, Goban2012}.  Here, we consider this phenomenon in a fully solid-state integrated photonics implementation consisting of a WGM resonator coupled to a QD with two Zeeman sublevels.

\section{Helicity-Sensitivity in Cavity QED -- A Linearized Master Equation Approach}

\begin{figure}[tb]
	\centering
	\includegraphics[scale=1.0]{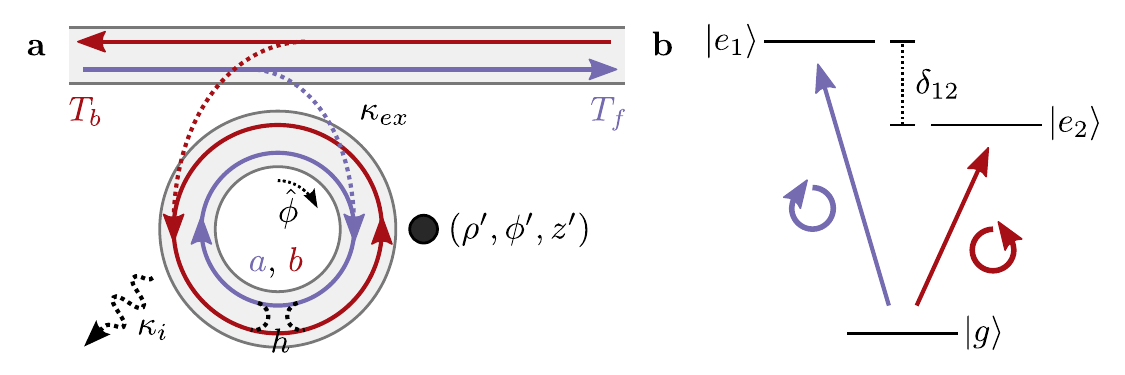}
	\caption{(a) Schematic of our non-reciprocal optical device, consisting of a waveguide evanescently coupled to a WGM cavity interacting with a QD at $(\rho',\phi',z')$. $\kappa_{\rm ex}$ is the coupling from the waveguide into the cavity and vice-versa, $\kappa_{\rm i}$ is the decay rate of the cavity modes, and $h$ is the backscattering in the cavity. $T_{\rm f}$ and $T_{\rm b}$ are the transmission for forward and backward propagating light in the waveguide respectfully. (b) Energy levels of the quantum dot. The two excited states couple to electric fields of opposite helicity.}
	\label{system}
\end{figure}

We construct a cavity QED model for an axisymmetric microcavity coupled to transitions sensitive to optical helicity.  The cavity is modeled as a single resonance of frequency $\omega_{\rm C}$, with two nominally degenerate counterpropagating modes. The system is excited by fields from either direction of frequency $\omega_{\rm p}$, assumed to be close to $\omega_{\rm C}$. Modes propagating forward through the waveguide couple only to one mode of the cavity (the CW mode), while backward propagating modes couple to the other (the CCW mode). The Hamiltonian for this system is
\beq
	\mathcal{H}_{\rm C} = \omega_{\rm C} \left( a^\dagger a + b^\dagger b \right) + h \left( a b^\dagger + b a^\dagger \right) + \mathcal{E}^*_{\rm p}e^{i\omega_{\rm p}t}o+\mathcal{E}_{\rm p} e^{-i\omega_{\rm p} t}o^\dagger
\eeq
where $a$ and $b$ are the bosonic annihilation operators for CW and CCW modes respectfully, $h$ is the intermode coupling due to backscattering, $\mathcal{E}_{\rm p}$ is the driving field amplitude, and $o$ is $a$ ($b$) when describing purely forward (backward) propagating light input into the waveguide. Shifting to a reference frame rotating at $\omega_{\rm p}$, the Hamiltonian becomes
\beq
	\mathcal{H}_{\rm C} = \Delta_{\rm C} \left( a^\dagger a + b^\dagger b \right) + h \left( a b^\dagger + b a^\dagger \right) + \mathcal{E}^*_p o+\mathcal{E}_p o^\dagger
\eeq
where $\Delta_{\rm C} = \omega_{\rm C}-\omega_{\rm p}$ is the cavity-probe detuning.

Since the $a$ and $b$ modes are counter-propagating, their respective electric field mode functions have opposite helicity $P$ at all positions, with magnitude $p = |P|$. This can be exploited to create non-reciprocal transmission at certain frequencies in the waveguide by having the cavity interact with a system sensitive to the helicity of the electric field. As in~\cite{Shen2011} and~\cite{Mi2011}, we consider a V-type three-level system consisting of a ground state, $\ket{g}$, and two excited states, $\ket{e_1}$ and $\ket{e_2}$ with transitions to each excited state driven by opposite electric field helicity~(Fig.~\ref{system}b). Such a system could be realized by a quantum dot with a Zeeman-like splitting (Sec.~\ref{sec:feasibility}). The QD is placed at a point $(\rho', \phi', z')$ within the evanescent field of the cavity where $\vect{E}$ is strong and possesses a high degree of helicity ($p \approx 1$).

Assuming the incident optical power is low and the QD is weakly excited, the two transitions can be approximated as independent (two separate 2-level systems instead of a 3-level system). The Hamiltonian for a QD with zero-splitting excitation energy of $\omega_{\rm QD}$ is
\begin{equation}
	\mathcal{H}_{\rm QD} = ( \omega_{\rm QD} + \delta_{12}/2 ) \sigma_1^+ \sigma_1^- + ( \omega_{\rm QD} - \delta_{12}/2 ) \sigma_2^+ \sigma_2^-
\end{equation}
where $\delta_{12}$ is the excited state splitting and $\sigma_1^-$ and $\sigma_2^-$ are annihilation operators for the excited states. For simplicity, we fix the QD energy to be resonant with the cavity frequency $\omega_{\rm C}$.

The light-matter interaction is described to lowest order by the dipole Hamiltonian
\beq
	\mathcal{H}_{\rm int} = - \hat{\mathbf{d}} \cdot \hat{\mathbf{E}}
\eeq
where $\hat{\mathbf{d}}$ and $\hat{\mathbf{E}}$ are the dipole and field operators. The electric field operator can be expressed in the helicity basis using Eqs.~\ref{posnegcw} and \ref{posnegccw} and the boson operators for the optical modes:
\beq
    \hat{\mathbf{E}}(\rho,\phi,z) = E_{\rm CW,+}\hat{e}_{+} a + E_{\rm CW,-}\hat{e}_{-} a + E_{\rm CCW,+}\hat{e}_{+} b + E_{\rm CCW,-}\hat{e}_{-} b + \textrm{c.c.}
\eeq
Since the excited state transitions of the QD are sensitive to electric field helicity, the only nonzero dipole matrix elements are $\bra{e_1}\hat{\vect{d}}\cdot\hat{e}_+\ket{g}$ and $\bra{e_2}\hat{\vect{d}}\cdot\hat{e}_-\ket{g}$. Assuming that these matrix elements have the same magnitude $d$ and the electric field modes obey the symmetries of WGMs, we can define a complex dipole coupling strength coefficient at the position of the QD
\begin{align}
	g(\vect{r}') &= -d|\mathbf{E}_M(\rho',z')|e^{iM\phi'} \nonumber \\
	&= g_0(\rho', z') e^{i\vartheta}
\end{align}
with $\vartheta$ the phase difference between $g$ and the mode coupling $h$. $g$ parameterizes a Jaynes-Cummings type Hamiltonian in the rotating wave approximation:
\begin{align}
	\mathcal{H}_{\rm int} =~&( g_+a\sigma_1^++g_+^*\sigma_1^-a^\dagger ) + ( g_-a\sigma_2^++g_-^*\sigma_2^-a^\dagger ) \nonumber\\
	& +~ ( g^*_-b\sigma_1^++g_-\sigma_1^-b^\dagger ) + ( g^*_+b\sigma_2^++g_+\sigma_2^-b^
\dagger ) 	\label{hint2}
\end{align}
where $g_\pm = g[(1 \pm p)/2]^{1/2}$.

We adopt a master equation approach and input-output formalism to account for waveguide coupling and system loss~\cite{Srinivasan2007.02}. Using the total Hamiltonian $\mathcal{H}= \mathcal{H}_{\rm C} + \mathcal{H}_{\rm QD} + \mathcal{H}_{\rm int}$, the time evolution of the density matrix $\rho$ for the system is given by the master equation
\begin{align}
	\dot{\rho} = -i[\mathcal{H},\rho]+2\kappa L(a)\rho + 2\kappa L(b)\rho + \gamma L(\sigma_1^-)\rho + \gamma L(\sigma_2^-)\rho
\end{align}
where $L(O)\rho = O\rho O^\dagger-\frac{1}{2}(O^\dagger O\rho-\rho O^\dagger O)$ is the Lindblad superoperator for operator $O$ acting on $\rho$, $\kappa$ is the sum of the waveguide-cavity coupling $\kappa_{\rm ex}$ and the cavity loss rate $\kappa_{\rm i}$, and $\gamma$ is the decay rate of the two excited states of the QD, assumed to be equal.

We seek a steady-state solution to the master equation. Assuming again that the QD is weakly excited by low input power, the fermionic QD operators can be treated as approximately bosonic $\left(\square{\sigma_1^-,\sigma_1^+} = \square{\sigma_2^-,\sigma_2^+} \simeq 1\right)$.  From the master equation, we obtain a set of linear equations for the time derivatives of operator expectation values which can then be solved directly for steady-state solutions of the linearized master equation.

We consider the observed optical transmission and reflection in two cases: light input in the forward ($a$) direction, and light input in the reverse ($b$) direction.  Using the input-output formalism \cite{Gardiner1985} for this linearized model, when light is input to the waveguide in the forward direction, the normalized forward transmission $T_{\rm f}$ and the normalized reflection $R_{\rm f}$ are
\begin{align}
	T_{\rm f} &= \absol{ i + 2 \kappa_{\rm ex} \left< a \right> / \mathcal{E}_{\rm p} }^2\\
	R_{\rm f} &= \absol{  2 \kappa_{\rm ex} \left< b \right> / \mathcal{E}_{\rm p} }^2
\end{align}
For light sent in the backward direction (input in the $b$ mode), the backward transmission $T_{\rm b}$ and reflection $R_{\rm b}$ are given by the same equations, but with $a$ and $b$ interchanged.  In the remainder of this paper, we study the transmission spectra as a function of the cavity-probe laser detuning $\Delta_{\rm C}$ found through the steady-state solutions of the linearized master equation. Parameters used chosen based on recent cavity QED analyses~\cite{Srinivasan2007.02} and experiments~\cite{Srinivasan2007.12}. The phase difference between $g$ and $h$, $\vartheta$, has little effect on the conclusions; it is chosen to be $\pi$/4 for simplicity.

\section{Optical Non-reciprocity in Axisymmetric Cavity QED}

\begin{figure}[tb]
	\centering
    \includegraphics[scale=1.0]{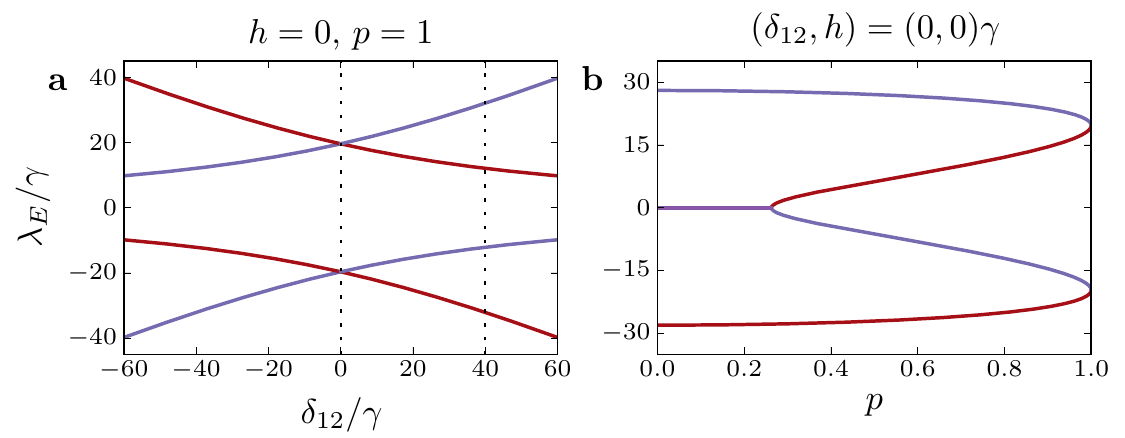}
	\caption{Exciton-polariton energy eigenvalues $\lambda_E$ of $\mathcal{H}$ as a function of (a) the splitting $\delta_{12}$ in the ideal case, and (b) the helicity $p$. Dashed lines indicate the conditions for the spectra in Fig~\ref{transplots}a-b and blue (red) indicates the forward (backward) sub-system. Other parameters are $(g_0,\kappa_{\rm i},\kappa_{\rm ex})=(20,3,5)\gamma$. The three eigenvalues for $p=0$ are characteristic of standard cavity QED with axisymetric resonators~\cite{Aoki2006,Dayan2008,Alton2011}. }
	\label{jceigsplots}
\end{figure}

\begin{figure}[tb]
	\centering
    \includegraphics[scale=1.0]{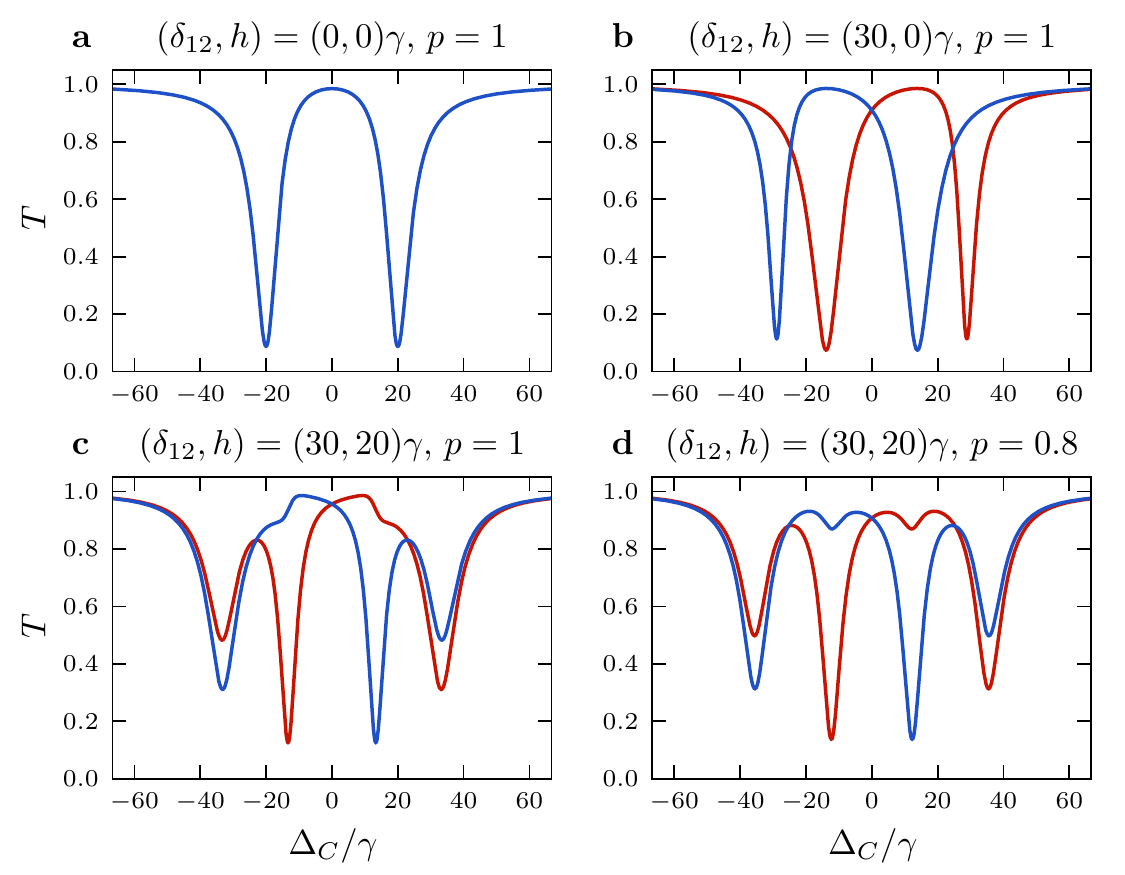}
	\caption{Forward (blue) and backward (red) transmission spectra in (a) the ideal case with zero excited-state splitting, (b) the ideal case with non-zero splitting, and (c-d) the non-ideal case with non-zero splitting. Other parameters are $(g_0,\kappa_{\rm i},\kappa_{\rm ex})=(20,3,5)\gamma$, $\vartheta=\pi/4$. }
	\label{transplots}
\end{figure}

To illustrate the directional asymmetry in this model, we first consider an idealized case with no backscattering $\paren{h=0}$ and perfect mode helicity $\paren{p=1}$. Without these forms of directional mixing, the forward and backward transmissions are fully decoupled; $T_{\rm f}$ and $T_{\rm b}$ depend only on the $a$ and $\sigma_1$ fields, or the $b$ and $\sigma_2$ fields, respectively. In analogy with the standard Jaynes-Cummings model, each propagating mode hybridizes with its coupled transition to form an independent pair of exciton-polariton eigenstates~(Fig~\ref{jceigsplots}a). When the excited state energy splitting $\delta_{12}$ is zero, the energy eigenvalues of the two polariton pairs are equal and the $T_{\rm f}$ and $T_{\rm b}$ spectra are characterized by identical dips in the transmission at detuning values of approximately $\pm g_0$ (Fig.~\ref{transplots}a). There is no difference between forward and backward transmission and the optical transmission is reciprocal.

For nonzero energy splitting $\delta_{12}$, the forward and backward transmission spectra are in general not equal~(Fig.~\ref{transplots}b). Increasing the spitting causes the cavity modes and relevant QD excited states to go off resonance: one polariton branch becomes more cavity-like and the other more exciton-like. Directional asymmetry in coupling leads to distinct spectral properties from the cavity QED model and non-reciprocal optical transmission.

The emergence of optical non-reciprocity in this model can be explained by time reversal symmetry breaking of the polarization-sensitive Zeeman excited states. If the excited states were time-reversal symmetric, the cavity QED model would be insensitive to direction ($T_{\rm f}=T_{\rm b}$). Since the time reversal symmetry is broken by a magnetic field generating non-zero $\delta_{12}$, the forward and backward directions are inequivalent and optical non-reciprocity is possible. Optical non-reciprocity can be further understood by examining the Hamiltonian under the time reversal operation. Although the magnetic field changes sign under time reversal, the energy splitting between excited states does not since the spins also reverse. The helicity of the WGM field changes sign, however, since the winding motion of the electric field inverts. As a result, $p \rightarrow -p$ and the interaction Hamiltonian is given by Eq. \ref{hint2}, except with $\sigma_1$ and $\sigma_2$ interchanged. As long as $\delta_{12}=0$, the Hamiltonians are equivalent and transmission is reciprocal. If there is an energy splitting, however, the symmetry is broken and the two directions are inequivalent.

When $|p|\neq1$ or $h\neq0$, directional input does not exclusively couple to just one excited state. Instead of two dips, $T_{\rm f}$ and $T_{\rm b}$ exhibit four dips at the eigenenergies of the four exciton-polaritons~(Fig.~\ref{transplots}c,d). Although still non-reciprocal for non-zero $\delta_{12}$, spectral differences are diminished, particularly near the exciton-like polariton. For zero helicity ($p=0$) when $\delta_{12}=0$, the model predicts three eigenvalues characteristic of an axisymmetric WGM coupled to a 2-level system, which would manifest as three dips in $T_{\rm f}$ and $T_{\rm b}$~\cite{Aoki2006,Dayan2008,Alton2011}. As $|p|$ increases, the cavity-like polariton in the eigenvalue spectrum divides into two branches for intermediate helicities before reaching the ideal directional degeneracy at $|p|=1$~(Fig.~\ref{jceigsplots}b).

\section{Single-Mode Optical Isolation from Directional Asymmetry}

The directional asymmetry shown in Fig.~\ref{transplots}b-d suggests that this system can function as a coherent optical diode~\cite{Hafezi2012}. The requirements for a non-reciprocal system to function as an optical isolator have been discussed recently~\cite{Jalas2013,Fan2012.01.06}. First, there must exist a mode of the waveguide for which the backward transmission is near zero while the forward transmission is near unity. This is satisfied here by tuning the parameters so that $T_f \simeq 1$ and $T_b \simeq 0$ at the same cavity-probe detuning. The second requirement is that transmission for all backward input modes must be blocked. This requirement is satisfied here by using a single-mode waveguide, as additional frequency modes not resonant with the cavity would be unblocked. This system can be classified as a single-mode optical diode; it provides optical isolation for a narrow-band of frequencies for a particular propagating TE or TM mode.

Here, we consider the system parameters that allow non-reciprocal isolator-like performance.  For ideal directional contrast (defined as $T_{\rm f}/T_{\rm b}$), the system parameters must be optimized for the highest forward transmission at the lowest possible value of the backward transmission.  Parameters that can be controlled experimentally include: $\Delta_{\rm C}$ by tuning the frequency of input light, $\delta_{12}$ by modifying an external magnetic field, and $\kappa_{\rm ex}$ by adjusting the distance between the waveguide and the cavity.  The remaining parameters, $g_0$, $\gamma$, $\kappa_{\rm i}$, $p$, and $h$ are considered fixed by the fabrication process and are treated as constants.  For this analysis, we seek to maximize $T_{\rm f}/T_{\rm b}$ in the three-dimensional parameter space~$(\Delta_{\rm C}, \delta_{12}, \kappa_{\rm ex})$.

\subsection{The Ideal Case}

Ideal optical diode behavior requires that for a particular light frequency $\omega_p$, $T_{\rm f} \approx 1$ and $T_{\rm b} = 0$. To obtain simple expressions for the parameters required for diode behavior, we first focus on the ideal case with no directional mixing, $h=0$ and $p=1$. The forward and backward transmission can be written analytically as
\begin{equation}
	T_{\rm f,b} = \left| 1-\frac{2\kappa_{\rm ex}[\gamma/2+i(\Delta_{\rm C} \pm \delta_{12}/2)]}{g_0^2+[\gamma/2+i(\Delta_{\rm C} \pm \delta_{12}/2)](\kappa_{\rm ex}+\kappa_{\rm i}+i\Delta_{\rm C})} \right|^2
	\label{idealtrans}
\end{equation}
where the positive (negative) signs correspond to the forward (backward) transmission direction.  By setting $T_{\rm b} = 0$ in Eq.~\ref{idealtrans}, restrictions on two of the controllable parameters, chosen arbitrarily to be $\delta_{12}$ and $\Delta_{\rm C}$, can be derived so that $T_{\rm b} = 0$:
\begin{align}
	\delta_{12} &= [\gamma-2(\kappa_{\rm ex}-\kappa_{\rm i})] \left( \frac{2g_0^2}{\gamma(\kappa_{\rm ex}-\kappa_{\rm i})}-1 \right) ^{1/2}
	\label{splittingcond}\\
	\Delta_{\rm C} &= -(\kappa_{\rm ex}-\kappa_{\rm i}) \left( \frac{2g_0^2}{\gamma(\kappa_{\rm ex}-\kappa_{\rm i})}-1 \right) ^{1/2}
	\label{detuningcond}
\end{align}
Simultaneouly satisfying these conditions puts two constraints on the system: $\kappa_{\rm ex} > \kappa_{\rm i}$ and $g_0^2 \geq \gamma(\kappa_{\rm ex}-\kappa_{\rm i})/2$. For allowed values of $\kappa_{\rm ex}$, we distinguish two regimes in which $T_{\rm b}=0$ is possible: for $\kappa_{\rm i} < \kappa_{\rm ex} < \kappa_{\rm i}+\gamma/2$, $T_{\rm b} =0$ occurs when $\Delta_{\rm C}$ matches the energy of the cavity-like polariton, and for $\kappa_{\rm ex} > \kappa_{\rm i} + \gamma/2$, it occurs when $\Delta_{\rm C}$ coincides with the QD-like polariton.

\begin{figure}[tb]
	\centering
    \includegraphics[scale=1.0]{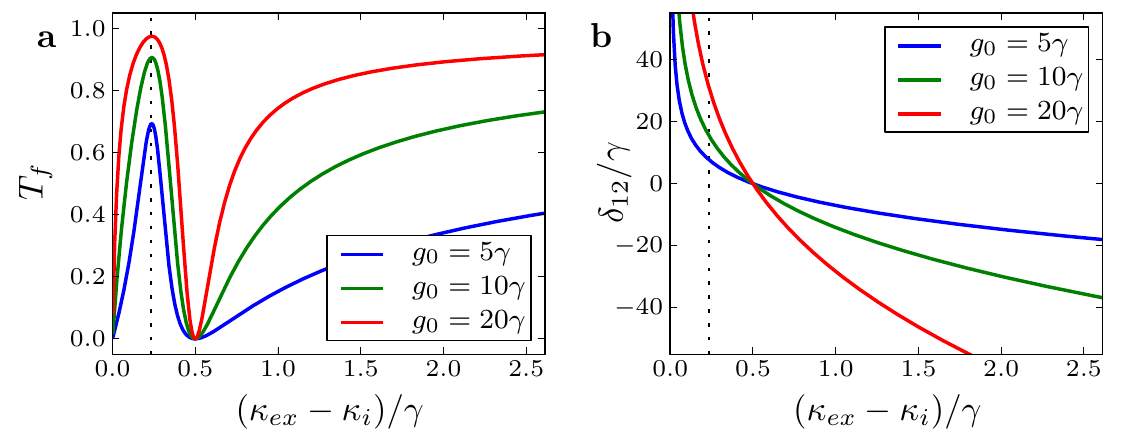}
	\caption{(a) The forward transmission with $\delta_{12}$ and $\Delta_{\rm C}$ set by Eq.~\ref{splittingcond}-\ref{detuningcond}. (b) The splitting $\delta_{12}$ set by Eq.~\ref{splittingcond} for several values of $g_0$. The dashed line shows the optimal value of $\kappa_{\rm ex}$ to maximize $T_{\rm f}$ for $g_0=20\gamma$. Other parameters are $(h,\kappa_{\rm i})=(0,5)\gamma$, $p=1$, and $\vartheta=\pi/4$.}
	\label{ideal0BTPlots}
\end{figure}

With the restrictions on $\delta_{12}$ and $\Delta_{\rm C}$ given by Eqs.~\ref{splittingcond}-\ref{detuningcond}, $T_{\rm f}$ can be optimized with the last controllable parameter, $\kappa_{\rm ex}$. By examining the forward transmission as a function of $\kappa_{\rm ex}$ with $\delta_{12}$ and $\Delta_{\rm C}$ set to enable zero backward transmission, plotted in Fig.~\ref{ideal0BTPlots}a, it can be seen that that high forward transmission is achievable when $\kappa_{\rm ex}$ is is in the regime $\kappa_{\rm i} < \kappa_{\rm ex} < \kappa_{\rm i}+\gamma/2$ even for a low value of the coupling strength $g_0$. Additionally, this regime requires a lower value of $\delta_{12}$ for high $T_{\rm f}$ than the the the regime $\kappa_{\rm ex} > \kappa_{\rm i}+\gamma/2$ as can be seen in Fig \ref{ideal0BTPlots}b. For the choice of parameters $(g_0,\kappa_{\rm i}) = (20, 5)\gamma$ and $\vartheta = \pi/4$, the optimum value of $\kappa_{\rm ex}$ is $5.2\gamma$, giving a forward transmission of 97.5\% with a required $\delta_{12}$ and $\Delta_{\rm C}$ of $30.3\gamma$ and $-13.8\gamma$ respectively.

Choosing to find restrictions on a different pairs of parameters other than $\delta_{12}$ and $\Delta_{\rm C}$ to satisfy $T_{\rm b} = 0$ results in the same conclusion. One can visualize a one-dimensional line in the three-dimensional parameter space spanned by $\kappa_{\rm ex}$, $\delta_{12}$, and $\Delta_{\rm C}$ which give $T_{\rm b} = 0$. The plots in Fig.~\ref{ideal0BTPlots}b can be interpreted as the projections of this line on the $\kappa_{\rm ex}$-$\delta_{12}$ plane for various values of $g_0$. Along the line there exists an optimal point at which $T_{\rm b}$ is maximum and the optical isolation is greatest.

\subsection{The Non-Ideal Case}

Having established that near-perfect single-mode optical isolation can be achieved with ideally tuned parameters, we briefly consider performance under more realistic, non-ideal conditions ($h \neq 0$, $p \neq \pm 1$). In Fig.~\ref{transplots}c-d, it can be seen that the forward transmission is impaired by mode mixing and non-unity helicity of the electric field, yet the spectra still display a high contrast $T_{\rm f}/T_{\rm b}$. Although the transmission cannot be written in a simple analytic form as Eq.~\ref{idealtrans}, we may use our results for the optimal values of $\delta_{12}$, $\Delta_{\rm C}$, and $\kappa_{\rm ex}$ from analysis of the ideal case as initial values in a numerical optimization of the the contrast in the non-ideal case.

Using numerical methods, we find that $T_{\rm b} = 0$ is still achievable along a line of values in the space spanned by $\kappa_{\rm ex}$, $\delta_{12}$, and $\Delta_{\rm C}$. Again, there exists an optimal point along this line at the contrast is greatest. In general, for increasing $h$ or decreasing $p$, the optimal value of $\kappa_{\rm ex}$ increases whereas the optimal value of $\delta_{12}$ changes little from the ideal case.

\begin{figure}[tb]
	\centering
	\includegraphics[scale=1.0]{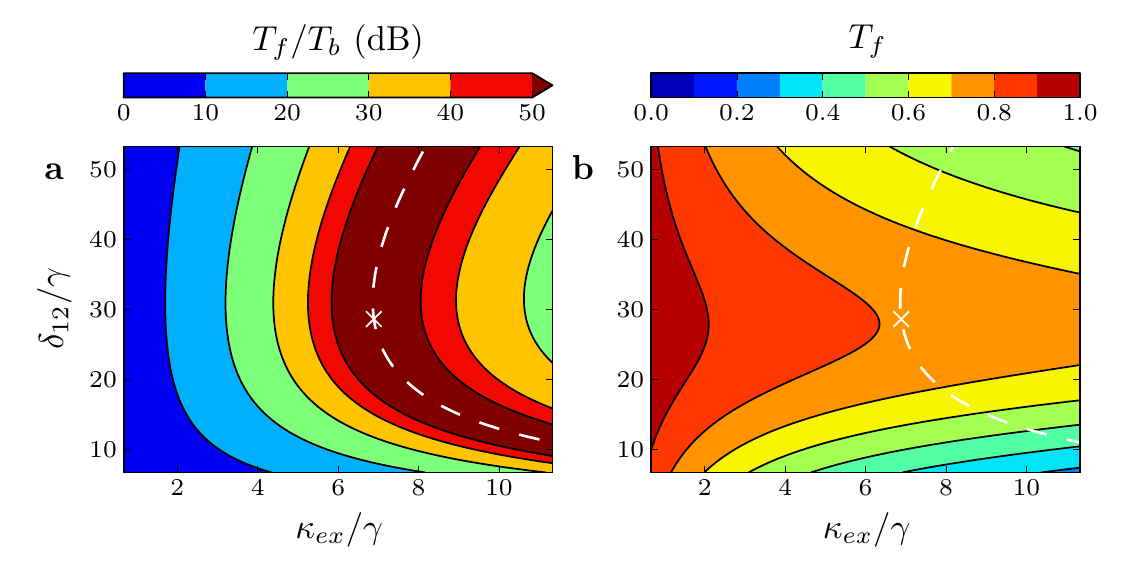}
	\caption{Contour plots of (a) the isolation contrast $T_{\rm f}/T_{\rm b}$ and (b) $T_{\rm f}$ for variable $\kappa_{\rm ex}$ and $\delta_{12}$ in the non-ideal case of $h = 20\gamma$, $p = 0.8$ and $\Delta_{\rm C}$ set to the cavity-like dip in the $T_{\rm b}$ spectrum. The dashed line indicates where $T_{\rm b} = 0$ may occur and the cross indicates the parameter location of optimal contrast.  Other parameters are $(g_0,\kappa_{\rm i}) = (20,5)\gamma$ and $\vartheta=\pi/4$.}
	\label{nonideal0BTContours}
\end{figure}

The isolation contrast is plotted in Fig.~\ref{nonideal0BTContours}a as a function of $\kappa_{\rm ex}$ and $\delta_{12}$, with $\Delta_{\rm C}$ set to the cavity-like dip in the backwards transmission spectrum. The projection of the line of values that enable $T_{\rm b} = 0$ in the $\kappa_{\rm ex}$-$\delta_{12}$ plane can be seen as well as the optimal point where the isolation contrast is greatest. If $\kappa_{\rm ex}$ and $\delta_{12}$ can be tuned to within about 30\% of their optimal values, isolation of over 30 dB can be achieved with a forward transmission of over 70\%~(Fig. \ref{nonideal0BTContours}b).

\section{Feasibility and Implementation}\label{sec:feasibility}

The formal conditions for optical isolation in the ideal ring resonator cavity QED system are $\kappa_{\rm ex} > \kappa_{\rm i}$ and $g_0^2 \geq \gamma(\kappa_{\rm ex}-\kappa_{\rm i})/2$. These translate to saying that critical waveguide-cavity coupling must be achievable and the light-matter coupling should be in a high cooperativity regime. Strong QD-cavity coupling is not required for optical isolation. Both requirements are achievable with current cavity QED methods~\cite{Srinivasan2007.02}. Early work with nanocrystal QDs demonstrated typical coupling $\hbar g > 10$ $\mu$eV in a WGM cavity and $\hbar\gamma \sim 6$ $\mu$eV~\cite{LeThomas2006}, and self-assembled QDs can have even more favorable parameters~\cite{Obrien2009}.  These numbers suggest that the required cavity parameters can be achieved with WGM resonators of quality $Q \gtrsim 10^5$, well within the state-of-the-art for lithographically-processed rings~\cite{Tien2011} and freestanding microresonators~\cite{Vahala2003}.

A more challenging requirement for realizing optical isolation in axisymmetric cavity QED is achieving helicity-sensitive eigenstates dominant over QD anisotropy~\cite{Htoon2009}. Additionally, the splitting between excited states must be on the order of the cavity-QD coupling. For typical QDs, a magnetic field on the order of 1~T would be necessary to overcome linear fine structure splitting so that the V-type three-level model is accurate and to achieve a sufficiently large splitting~\cite{Kuther1998}. Although sufficient fields can easily be applied externally in the lab, a miniaturized integrated photonics approach is less straightforward. Obtaining the required Zeeman splitting may be feasible using diluted magnetic semiconductor QDs~\cite{Beaulac2008, Pandey2012}, which could provide optically-induced splitting for excitons of greater than 10~T~\cite{Beaulac2009}.

\section{Conclusion}

We have analyzed direction-dependent transmission through a waveguide coupled to an axisymmetric resonator interacting with a helicity-sensitive quantum dot. Lifting the excited-state degeneracy with a Zeeman-type splitting induces non-reciprocal transmission. With properly tuned parameters, optical isolation can be achieved for light in a coherent single-photon regime, which could be useful for coherent optical diodes and phase shifting in quantum photonics.

\section*{Acknowledgements}

This work was funded by the Institute for Sustainability and Energy at Northwestern.  E.~J.~L. is part of the NSF IGERT (DGE-080168), and  N.~P.~S. is an Alfred P. Sloan Research Fellow.

\bibliography{Lenferink2014}

\begin{thebibliography}{10}
\newcommand{\enquote}[1]{``#1''}

\bibitem{Potton2004}
R.~J. Potton, \enquote{Reciprocity in optics,} Rep. Prog. Phys. \textbf{67},
  717 (2004).

\bibitem{Politi2008}
A.~Politi, M.~J. Cryan, J.~G. Rarity, S.~Yu, and J.~L. O'Brien,
  \enquote{Silica-on-silicon waveguide quantum circuits,} Science \textbf{320},
  646--649 (2008).

\bibitem{Politi2009}
A.~Politi, J.~C.~F. Matthews, and J.~L. O'Brien, \enquote{Shor's quantum
  factoring algorithm on a photonic chip,} Science \textbf{325}, 1221 (2009).

\bibitem{Shadbolt2011}
P.~J. Shadbolt, M.~R. Verde, A.~Peruzzo, A.~Politi, A.~Laing, M.~Lobino,
  J.~C.~F. Matthews, M.~G. Thompson, and J.~L. O'Brien, \enquote{Generating,
  manipulating and measuring entanglement and mixture with a reconfigurable
  photonic circuit,} Nature Photonics \textbf{6}, 45--49 (2012).

\bibitem{Obrien2009}
J.~L. O'Brien, A.~Furusawa, and J.~Vu\v{c}kovi\'{c}, \enquote{Photonic quantum
  technologies,} Nature Photon. \textbf{3}, 687--695 (2009).

\bibitem{Kilper2012}
D.~Kilper, K.~Guan, K.~Hinton, and R.~Ayre, \enquote{Energy challenges in
  current and future optical transmission networks,} Proceedings of the IEEE
  \textbf{100}, 1168--1187 (2012).

\bibitem{Koch2010}
J.~Koch, A.~A. Houck, K.~L. Hur, and S.~M. Girvin,
  \enquote{Time-reversal-symmetry breaking in circuit-{QED}-based photon
  lattices,} Phys. Rev. A \textbf{82}, 043811 (2010).

\bibitem{Wang2008}
Z.~Wang, Y.~D. Chong, J.~D. Joannopoulos, and M.~Solja\ifmmode \check{c}\else
  \v{c}\fi{}i\ifmmode~\acute{c}\else \'{c}\fi{}, \enquote{Reflection-free
  one-way edge modes in a gyromagnetic photonic crystal,} Phys. Rev. Lett.
  \textbf{100}, 013905 (2008).

\bibitem{Haldane2008}
F.~D.~M. Haldane and S.~Raghu, \enquote{Possible realization of directional
  optical waveguides in photonic crystals with broken time-reversal symmetry,}
  Phys. Rev. Lett. \textbf{100}, 013904 (2008).

\bibitem{Hafezi2011}
M.~Hafezi, E.~A. Demler, M.~D. Lukin, and J.~M. Taylor, \enquote{Robust optical
  delay lines with topological protection,} Nature Phys. \textbf{7}, 907--912
  (2011).

\bibitem{Hafezi2012}
M.~Hafezi and P.~Rabl, \enquote{Optomechanically induced non-reciprocity in
  microring resonators,} Opt. Express \textbf{20}, 7672--7684 (2012).

\bibitem{Bi2011}
L.~Bi, J.~Hu, P.~Jiang, D.~H. Kim, G.~F. Dionne, L.~C. Kimerling, and C.~A.
  Ross, \enquote{On-chip optical isolation in monolithically integrated
  non-reciprocal optical resonators,} Nature Photonics \textbf{5}, 758--762
  (2011).

\bibitem{Soljacic2003}
M.~Solja\v{c}i\'{c}, C.~Luo, J.~D. Joannopoulos, and S.~Fan, \enquote{Nonlinear
  photonic crystal microdevices for optical integration,} Opt. Lett.
  \textbf{28}, 637--639 (2003).

\bibitem{Fan2012}
L.~Fan, J.~Wang, L.~T. Varghese, H.~Shen, B.~Niu, Y.~Xuan, A.~M. Weiner, and
  M.~Qi, \enquote{An all-silicon passive optical diode,} Science \textbf{335},
  447--450 (2012).

\bibitem{Feng2011}
L.~Feng, M.~Ayache, J.~Huang, Y.-L. Xu, M.-H. Lu, Y.-F. Chen, Y.~Fainman, and
  A.~Scherer, \enquote{Nonreciprocal light propagation in a silicon photonic
  circuit.} Science \textbf{333}, 729--33 (2011).

\bibitem{Trowbridge2011}
C.~J. Trowbridge, B.~M. Norman, J.~Stephens, A.~C. Gossard, D.~D. Awschalom,
  and V.~Sih, \enquote{Electron spin polarization-based integrated photonic
  devices,} Optics Express \textbf{19}, 14845--14851 (2011).

\bibitem{Shen2011}
Y.~Shen, M.~Bradford, and J.~T. Shen, \enquote{Single-photon diode by
  exploiting the photon polarization in a waveguide,} Phys. Rev. Lett.
  \textbf{107}, 173902 (2011).

\bibitem{Mi2011}
X.~W. Mi, J.~X. Bai, D.~J. Li, and H.~P. Zhao, \enquote{Coupling to a microdisk
  cavity containing a three-level quantum-dot with two orthogonal modes,} Opt.
  Commun. \textbf{284}, 2937--2942 (2011).

\bibitem{Manipatruni2009}
S.~Manipatruni, J.~T. Robinson, and M.~Lipson, \enquote{Optical nonreciprocity
  in optomechanical structures,} Phys. Rev. Lett. \textbf{102}, 213903 (2009).

\bibitem{Lacroute2012}
C.~Lacro\^ute, K.~S. Choi, A.~Goban, D.~J. Alton, D.~Ding, N.~P. Stern, and
  H.~J. Kimble, \enquote{A state-insensitive, compensated nanofiber trap,} New
  J. Phys. \textbf{14}, 023056 (2012).

\bibitem{Goban2012}
A.~Goban, K.~S. Choi, D.~J. Alton, D.~Ding, C.~Lacro\^ute, M.~Pototschnig,
  T.~Thiele, N.~P. Stern, and H.~J. Kimble, \enquote{Demonstration of a
  state-insensitive, compensated nanofiber trap,} Phys. Rev. Lett.
  \textbf{109}, 033603 (2012).

\bibitem{Junge2013}
C.~Junge, D.~O'Shea, J.~Volz, and A.~Rauschenbeutel, \enquote{Strong coupling
  between single atoms and nontransversal photons,} Phys. Rev. Lett.
  \textbf{110}, 213604 (2013).

\bibitem{Aoki2006}
T.~Aoki, B.~Dayan, E.~Wilcut, W.~P. Bowen, A.~S. Parkins, T.~J. Kippenberg,
  K.~J. Vahala, and H.~J. Kimble, \enquote{Observation of strong coupling
  between one atom and a monolithic microresonator,} Nature \textbf{443},
  671--674 (2006).

\bibitem{Alton2011}
D.~J. Alton, N.~P. Stern, T.~Aoki, H.~Lee, E.~Ostby, K.~J. Vahala, and H.~J.
  Kimble, \enquote{Strong interactions of single atoms and photons near a
  dielectric boundary,} Nature Phys. \textbf{7}, 159--165 (2011).

\bibitem{Snyder1983}
A.~W. Snyder and J.~Love, \emph{Optical waveguide theory}, vol. 190 (Springer,
  1983).

\bibitem{Shen2009}
J.-T. Shen and S.~Fan, \enquote{Theory of single-photon transport in a
  single-mode waveguide. ii. coupling to a whispering-gallery resonator
  containing a two-level atom,} Phys. Rev. A \textbf{79}, 023838 (2009).

\bibitem{Oxborrow2007}
M.~Oxborrow, \enquote{How to simulate the whispering-gallery modes of
  dielectric microresonators in {FEMLAB/COMSOL},} in \enquote{Lasers and
  Applications in Science and Engineering,}  (International Society for Optics
  and Photonics, 2007), pp. 64520J--64520J.

\bibitem{Srinivasan2007.02}
K.~Srinivasan and O.~Painter, \enquote{Mode coupling and cavity{--}quantum-dot
  interactions in a fiber-coupled microdisk cavity,} Phys. Rev. A \textbf{75},
  023814 (2007).

\bibitem{Gardiner1985}
C.~W. Gardiner and M.~J. Collett, \enquote{Input and output in damped quantum
  systems: Quantum stochastic differential equations and the master equation,}
  Phys. Rev. A \textbf{31}, 3761 (1985).

\bibitem{Srinivasan2007.12}
K.~Srinivasan and O.~Painter, \enquote{Linear and nonlinear optical
  spectroscopy of a strongly coupled microdisk{--}quantum dot system,} Nature
  \textbf{450}, 862--865 (2007).

\bibitem{Dayan2008}
B.~Dayan, A.~S. Parkins, T.~Aoki, E.~P. Ostby, K.~J. Vahala, and H.~J. Kimble,
  \enquote{A photon turnstile dynamically regulated by one atom,} Science
  \textbf{319}, 1062--1065 (2008).

\bibitem{Jalas2013}
D.~Jalas, A.~Petrov, M.~Eich, W.~Freude, S.~Fan, Z.~Yu, R.~Baets, M.~Popovic,
  A.~Melloni, J.~D. Joannopoulos, M.~Vanwolleghem, C.~R. Doerr, and H.~Renner,
  \enquote{What is--and what is not--an optical isolator,} Nature Photonics
  \textbf{7}, 579--582 (2013).

\bibitem{Fan2012.01.06}
S.~Fan, R.~Baets, A.~Petrov, Z.~Yu, J.~D. Joannopoulos, W.~Freude, A.~Melloni,
  M.~Popovi{\'c}, M.~Vanwolleghem, D.~Jalas \emph{et~al.}, \enquote{Comment on
  {``}nonreciprocal light propagation in a silicon photonic circuit{''},}
  Science \textbf{335}, 38--38 (2012).

\bibitem{LeThomas2006}
N.~Le~Thomas, U.~Woggon, O.~Sch\"ops, M.~V. Artemyev, M.~Kazes, and U.~Banin,
  \enquote{Cavity {QED} with semiconductor nanocrystals,} Nano Letters
  \textbf{6}, 557--561 (2006).

\bibitem{Tien2011}
M.-C. Tien, J.~F. Bauters, M.~J.~R. Heck, D.~T. Spencer, D.~J. Blumenthal, and
  J.~E. Bowers, \enquote{Ultra-high quality factor planar {S}i$_3${N}$_4$ ring
  resonators on {S}i substrates,} Opt. Express \textbf{19}, 13551--13556
  (2011).

\bibitem{Vahala2003}
K.~J. Vahala, \enquote{Optical microcavities,} Nature \textbf{424}, 839--846
  (2003).

\bibitem{Htoon2009}
H.~Htoon, S.~A. Crooker, M.~Furis, S.~Jeong, A.~L. Efros, and V.~I. Klimov,
  \enquote{Anomalous circular polarization of photoluminescence spectra of
  individual {C}d{S}e nanocrystals in an applied magnetic field,} Phys. Rev.
  Lett. \textbf{102}, 017402 (2009).

\bibitem{Kuther1998}
A.~Kuther, M.~Bayer, A.~Forchel, A.~Gorbunov, V.~B. Timofeev, F.~Sch\"afer, and
  J.~P. Reithmaier, \enquote{Zeeman splitting of excitons and biexcitons in
  single {I}n$_{0.60}${G}a$_{0.40}${A}s/{G}a{A}s self-assembled quantum dots,}
  Phys. Rev. B \textbf{58}, 7508--7511 (1998).

\bibitem{Beaulac2008}
R.~Beaulac, P.~I. Archer, S.~T. Ochsenbein, and D.~R. Gamelin,
  \enquote{Mn$^{2+}$-doped {C}d{S}e quantum dots: {N}ew inorganic materials for
  spin-electronics and spin-photonics,} Adv. Func. Mater. \textbf{18},
  3873--3891 (2008).

\bibitem{Pandey2012}
A.~Pandey, S.~Brovelli, R.~Viswanatha, L.~Li, J.~M. Pietryga, V.~I. Klimov, and
  S.~Crooker, \enquote{Long-lived photoinduced magnetization in copper-doped
  {Z}n{S}e-{C}d{S}e core-shell nanocrystals,} Nat Nano \textbf{7}, 792--797
  (2012).

\bibitem{Beaulac2009}
R.~Beaulac, L.~Schneider, P.~I. Archer, G.~Bacher, and D.~R. Gamelin,
  \enquote{Light-induced spontaneous magnetization in doped colloidal quantum
  dots,} Science \textbf{325}, 973--976 (2009).

\end{thebibliography}

\end{document}